\documentclass[seceq,preprint]{ptptex}
\makeatletter
\def\bfm#1{\mbox{\boldmath $#1$}}

\def\vereq#1#2{\lower3pt\vbox{\baselineskip1.5pt \lineskip1.5pt
\ialign{$\m@th#1\hfill##\hfil$\crcr#2\crcr\sim\crcr}}}

\preprintnumber{\large YITP-04-8}
\makeatother
\usepackage[dvips]{color,graphicx}

\title{Improved Schwinger-Dyson approach\\
to pairing phenomena and QCD phase diagram%
\footnote{Poster talk presented by the author 
  for the Workshop on ``{\it Finite Density QCD at Nara}'',
  Nara, Japan, 10-12 July 2003.}
}
\author{
Hiroaki \textsc{Abuki}\footnote{\tt E-mail:abuki@yukawa.kyoto-u.ac.jp}%
}

\inst{
Yukawa Institute for Theoretical Physics,~Kyoto~606-8502,~Japan\\
}
\abst{
Phase structure and phase transitions in dense QCD are studied using 
  the Cornwall-Jackiw-Tomboulis (CJT) potential in the improved ladder 
  approximation.
The gap function, the condensation energy and the structure of Cooper 
  pairs are investigated at finite temperature and density.
Due to strong coupling effects at low densities, the gap, critical
  temperature and their ratio deviate from the weak coupling values. 
It is shown that the internal structure of Cooper pairs are robust 
  against the thermal effects, despite the fact that the pairs are 
  strongly correlated near the critical temperature.
Also, the effect of the strange quark mass, $m_s$ on the phase diagram is 
  examined using a simple kinematical criterion. 
We discuss the behavior of the unlocking line, on which the CFL turns
into the 2SC, through the variation of $m_s$.
}
\begin{document}
\newcommand{\beq}{\begin{eqnarray}}
\newcommand{\eeq}{\end{eqnarray}}
\newcommand{\e}{\epsilon}
\newcommand{\ds}[1]{#1 \hspace{-5.2pt}/} 
\newcommand{\Ds}[1]{#1 \hspace{-1ex}/} 
\newcommand{\isum}{\int\hspace{-14.0pt}\sum} 
\newcommand{\isums}{\int\hspace{-10.5pt}\sum} 
\newcommand{\bra}[1]{\left\langle\left. #1 \right\vert\right.}
\newcommand{\ket}[1]{\left.\left\vert #1 \right. \right\rangle}
\newcommand{\braket}[1]{\left\langle #1\right\rangle}
\maketitle
\section{Introduction}
The color superconducting phases at high baryon density have attracted 
  much interest in high density QCD.
In particular, the two types of the pairing pattern, namely color-flavor
  locked (CFL) pairing and iso-scalar (2SC) pairing have been studied by
  many authors and rich physics in these phases has
  been revealed so far.\cite{Review} 

In the chiral limit, 
  weak coupling analyses of the gap equations reveal
  the CFL dominance over the 2SC at zero temperature 
  and the coincidence of critical temperatures of transitions 
  to the QGP phase from the 2SC state and from the CFL state. 
The former is attributed to a competition between the number of degrees
  of freedom participating in the pairing correlation and the
  sizes of the gaps. 
The latter is due to two facts:~The rapid vanishing of the pairing in
  the color sextet channel as the critical temperature $T_c$ is
  approached, and the disappearance of the nonlinearity of the gap
  equation in the diquark condensate near $T_c$. 
However, in the low density regime, the sizes of gaps are modified
  from the weak coupling values owing to various strong coupling
  effects.\cite{itakura}
Also, the pairing in the color symmetric channel may become relevant at
  low densities, although it is suppressed by one coupling constant
  relative to the anti-triplet channel in the weak coupling
  regime. 

We examine in this article the Schwinger-Dyson (SD) approach
  \cite{Rischke-Review} in the improved ladder approximation to the
  pairing phenomena at finite
  temperature and density. 
This approach, the SD equation in the Landau gauge with an improved
  running coupling constant, can reproduce the reasonable vacuum physics
  related to the chiral symmetry breaking 
  as well as the perturbative result for the gap in the weak coupling
  regime.
Various strong coupling effects manifest themselves for realistic
  densities in this approach.\cite{itakura}
More complete and detailed analysis was done in Ref.~\cite{abuki}.

\section{Gaps, correlation and thermodynamic grand potential}
The gap equation is nothing but the self-consistency condition for
the proper self energy $\hat{\Delta}_\pm(p_0,p)$ in the Nambu-Gor'kov
base $\Psi=(q, i\gamma^2\gamma^0\bar{q}^t)$
\beq
  \bfm{\Sigma}(p_0,\bfm{p})%
  &=&ig^2\int\!\!\!\frac{d^4q}{(2\pi)^4} {\bfm{\Gamma}}_a^\mu\,%
     \bfm{S}(q_0,\bfm{q})\bfm{\Gamma}_b^\nu\,%
     D^{ab}_{\mu\nu}(q_0-p_0,\bfm{q}-\bfm{p}).
\eeq
Here, $D^{ab}_{\mu\nu}$ is the in-medium gluon propagator, $\bfm{S}$
is the full quark propagator, and
$\bfm{\Gamma}^a_\mu=\gamma_\mu\lambda^a/2$ is the quark-gluon vertex. 
The gap matrix is defined by the off-diagonal component of the self
energy $\bfm{\Sigma}$. The 2SC ansatz for the gap matrix is
\beq
  \hat{\Delta}^{\rm 2SC}_\pm(p_0,\bfm{p})%
  =(\tau_2\times\lambda_2)^{ab}_{ij}\Delta_\pm(p_0,\bfm{p}).
\eeq
$\Delta_{+(-)}$ is the positive (negative) energy component of the gap
matrix
$\Delta(p_0,\bfm{p})=(\Delta_-+\Delta_+)%
-(\Delta_+-\Delta_-)\gamma^0\bfm{\gamma}\cdot\hat{p}$. 
The color-flavor locked gap matrix is 
\beq
    \hat{\Delta}^{\rm CFL}_\pm(p_0,p)%
    &=&\frac{1}{3}\delta^a_i\delta^b_j\Delta^1_{\pm}(p_0,\bfm{p})%
        +\left(\delta^a_j\delta^b_i-\frac{1}{3}\delta^a_i\delta^b_j%
	\right)\Delta^8_{\pm}(p_0,\bfm{p}).
\eeq
Index $(a,b,\cdots)$ and $(i,j,\cdots)$ represents color
and flavor, respectively.
We use the quasi-static approximation of the hard dense loop
propagator in the Landau gauge\cite{itakura,abuki}.
Following Ref.~\cite{itakura}, we replace the
coupling $g^2$ with the momentum dependent effective coupling
$\bar{g}^2(p,q)$ using the Higashijima-Miransky
prescription.
\beq
   \bar{g}^2(p,q)%
   =\frac{16\pi^2}{\beta_0}\frac{1}{\ln((p_{\rm
     max}^2+p_c^2)/\Lambda^2)},\quad{p_{\rm max}}={\rm max}(p,q),
\eeq
where $\beta_0=(11N_c-2N_f)/3$, and $p_c^2$ plays the role of an
infrared regulator. We adopted $\Lambda=400$ MeV and
$p_c^2=1.5\Lambda^2$ for our numerical calculations.
The correlation function describes the off-diagonal long range order in
the system:
\beq
   \phi_{+}(q)%
   =\tanh\left({\sqrt{(q-\mu)^2+\Delta_+^2(q)}}%
   \Big/{2T}\right)\frac{\Delta_+(q)}%
   {2\sqrt{(q-\mu)^2+\Delta_+^2(q)}}.\label{eq:lkj}
\eeq
$\Delta(q)$ is the gap for the quasi-quark having momentum
$q=|\bfm{q}|$. $\phi_+(p)$ are defined by
$
\langle {\rm 2SC}|a^a_{i}(p)a^b_{j}(-p)|{\rm 2SC}\rangle%
=(\tau_2\lambda_2)^{ab}_{ij}\phi_+(q)%
$, where $a$ is the annihilation operator for quark.
We have omitted the helicity indices here. We can calculate the
coherence length from the correlation function
which characterizes the size of the quark Cooper pair.\cite{itakura,abuki}
Using the method developed by Cornwall, Jackiw and Tomboulis\cite{CJT}, 
the thermodynamic potential par unit volume relative to the normal Fermi
gas up to 2-loop order is given
by 
$
\delta\Omega(\mu,T)%
=\frac{T}{2V}\left\{{\rm
 TrLog}\left[\bfm{S}\bfm{S}_0^{-1}\right]-\frac{1}{2}{\rm Tr}%
 \big[\bfm{S}\bfm{\Sigma}\big]\right\}%
$. $\bfm{S}_0$ is bare quark propagator in the Fermi sea.
Note that this expression is only valid at the stationary point
determined by solving the gap equation for $\bfm{\Sigma}$.

\section{Numerical results and discussion}
\begin{figure}[tbp]
\centerline{	
\includegraphics[width=0.48\textwidth]{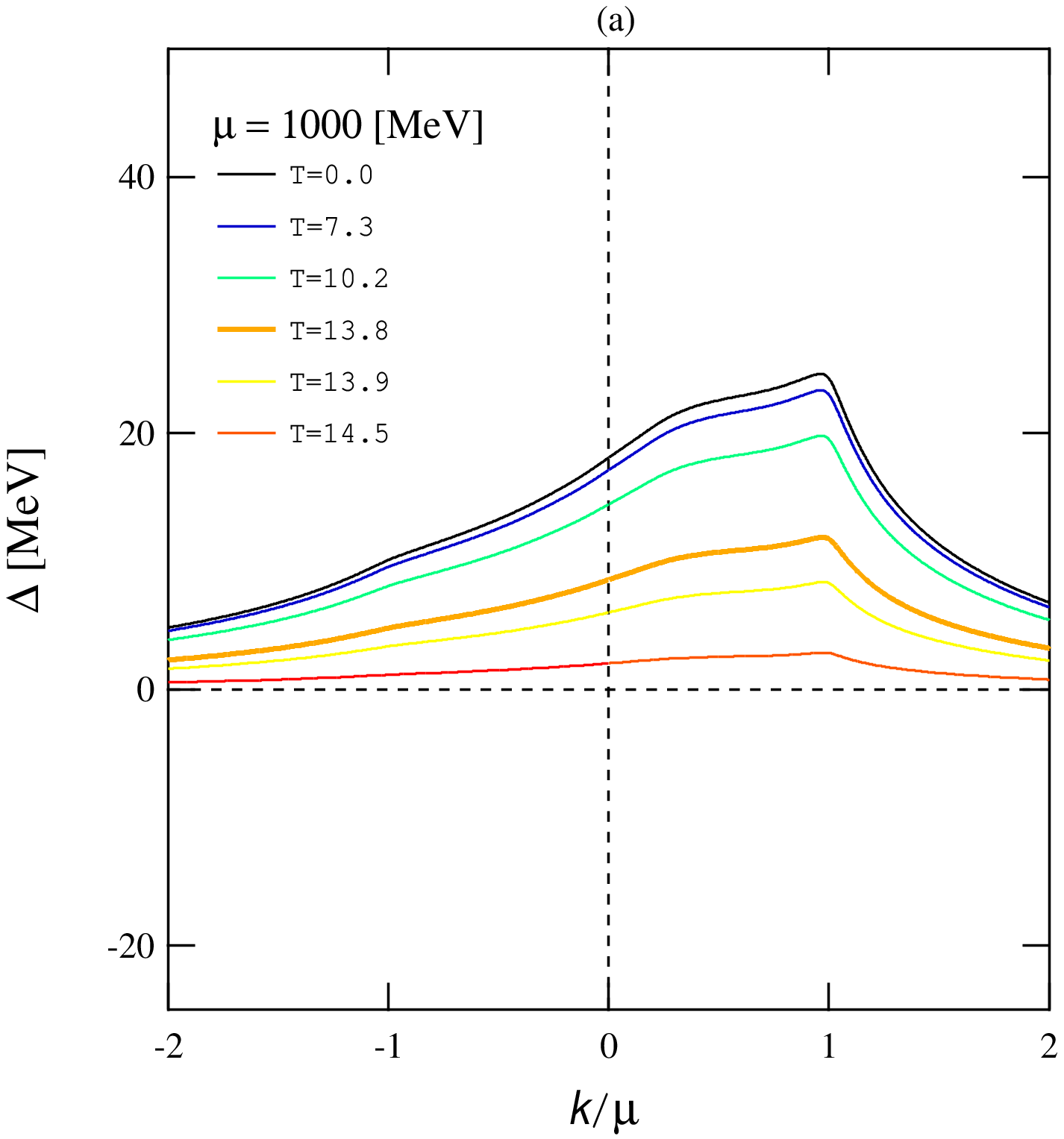}
\includegraphics[width=0.48\textwidth]{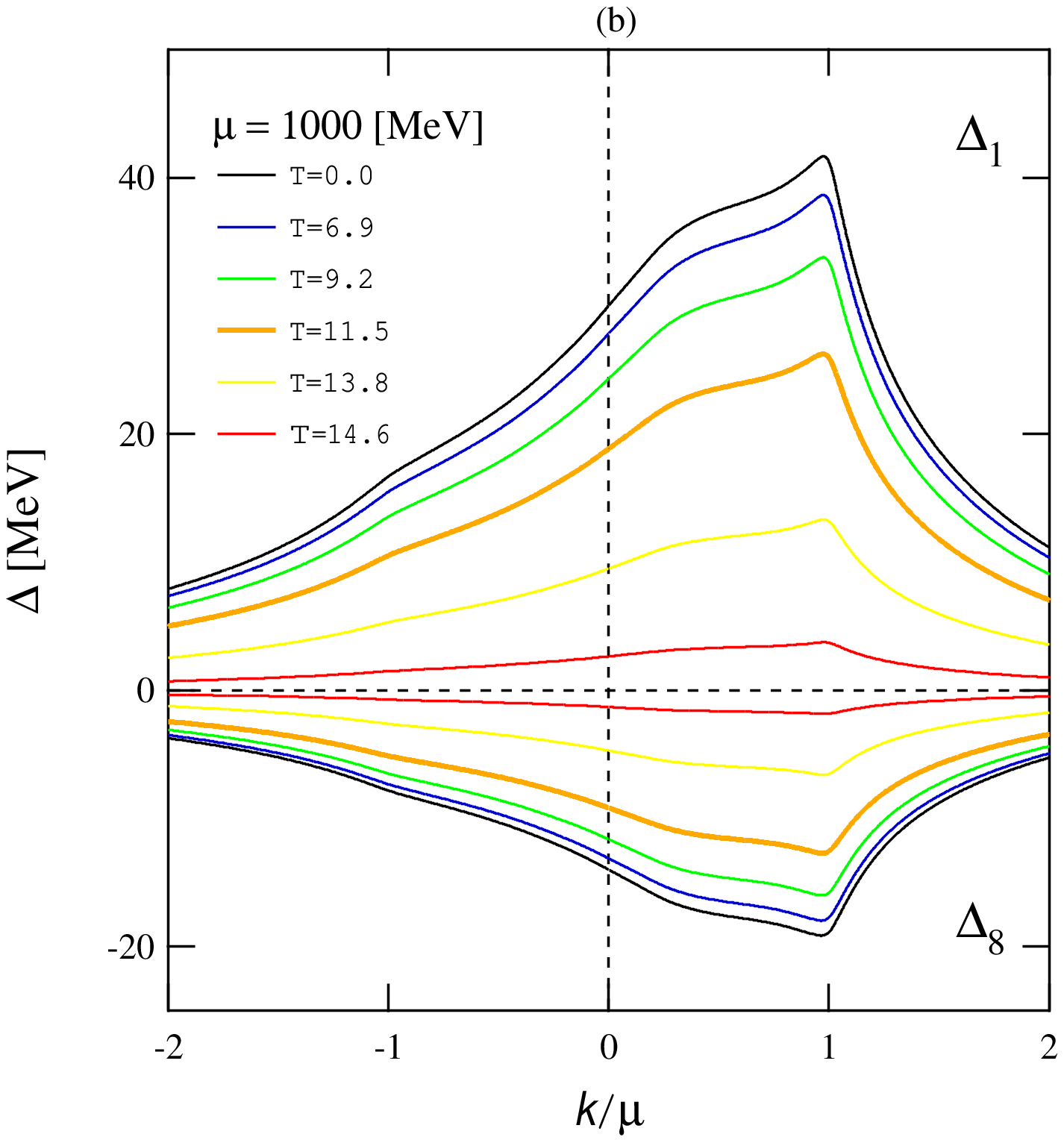}
}
\centerline{	
\includegraphics[width=0.48\textwidth]{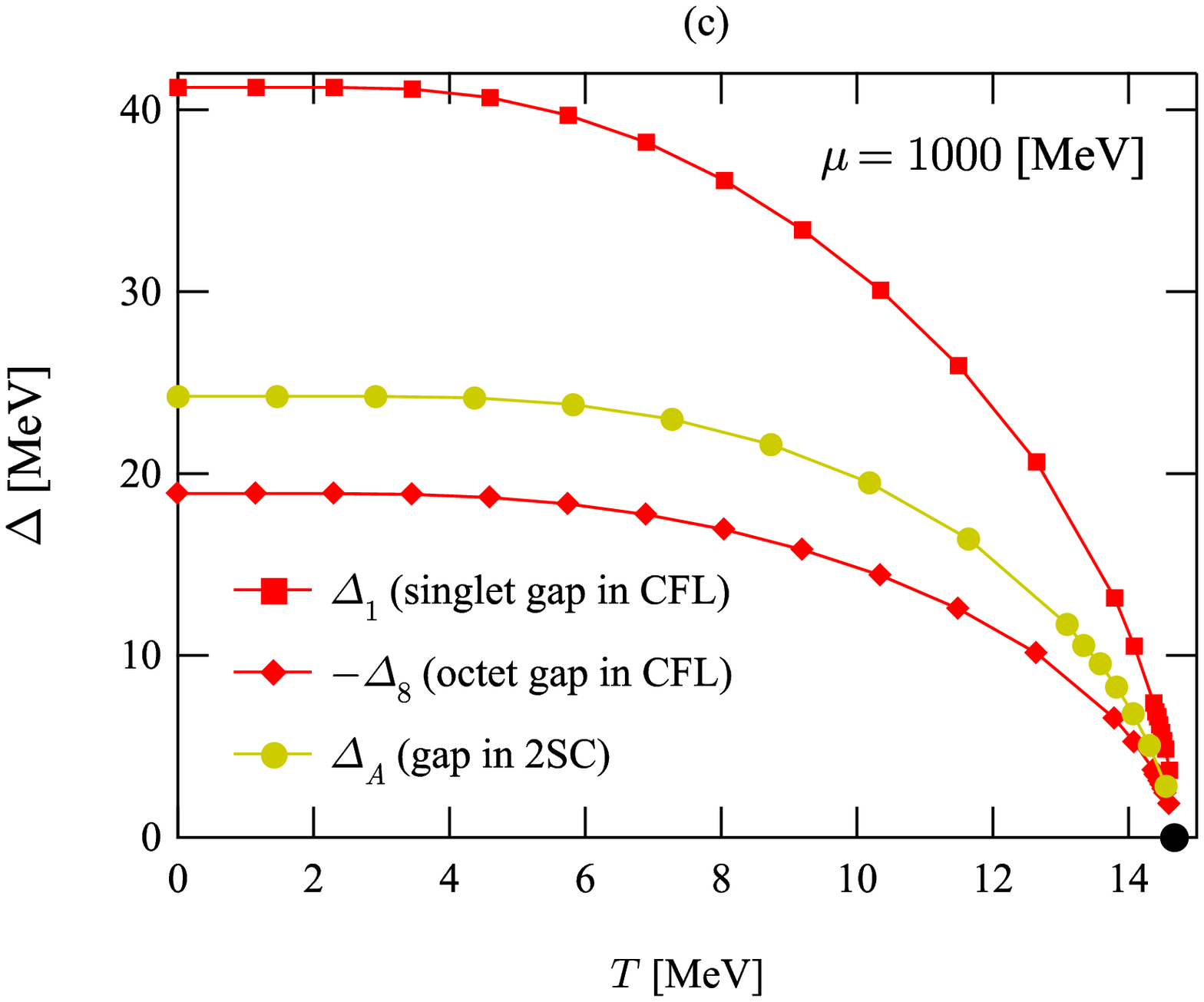}
\includegraphics[width=0.48\textwidth]{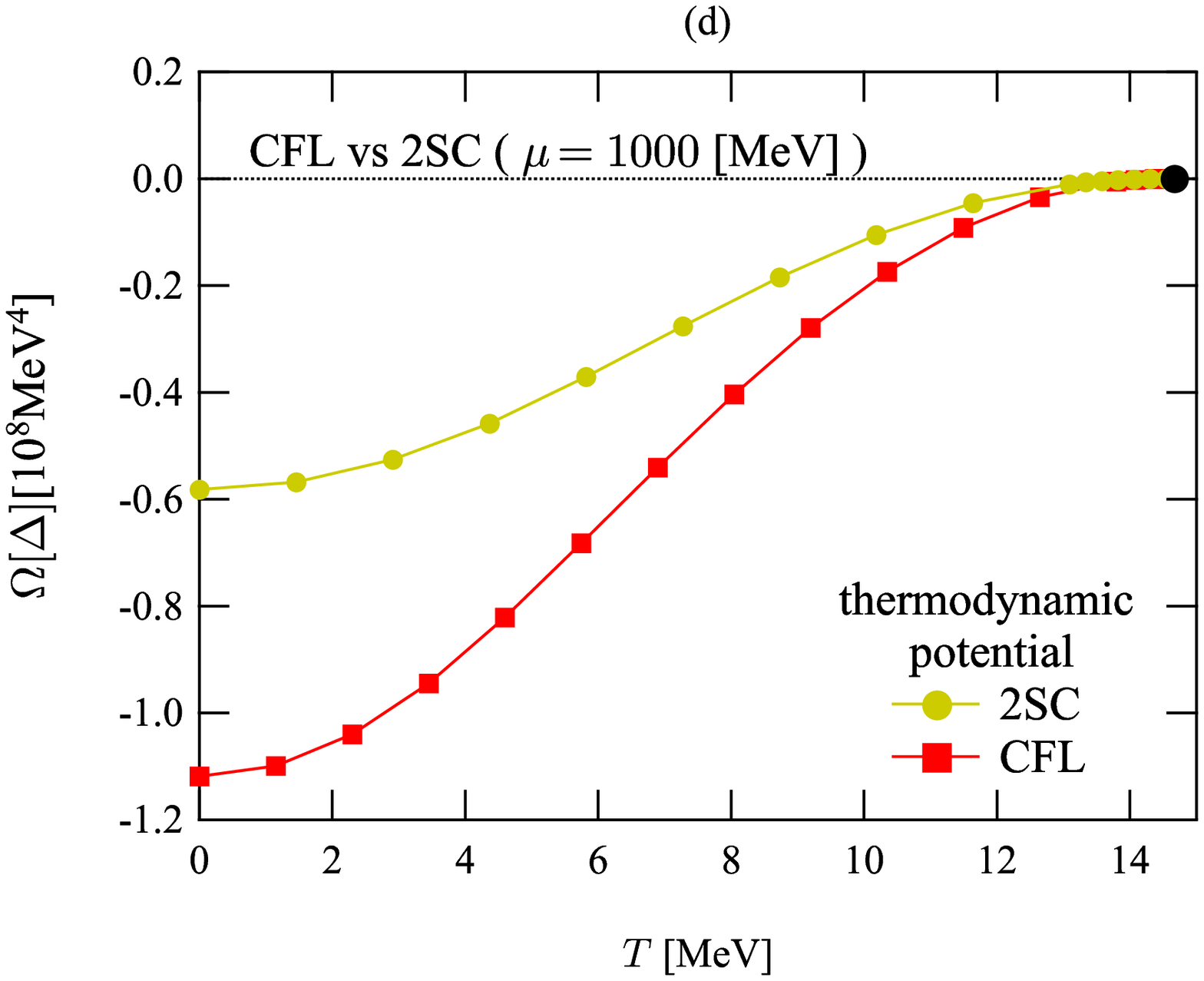}
}
\caption[]{
 (a) Gap functions at $\mu=1000$MeV for the 2SC state.
 The values $(-|k|, \Delta_-(|k|))$ are plotted in left half of the
 figure. 
 (b) Gap functions for the CFL state.
 (c) The temperature dependence of the gaps at the Fermi surface.
 (d) The condensation energy as function of temperature.}
\label{fig:0}  
\end{figure}

\begin{figure}[tbp]
\centerline{	
\includegraphics[width=0.48\textwidth]{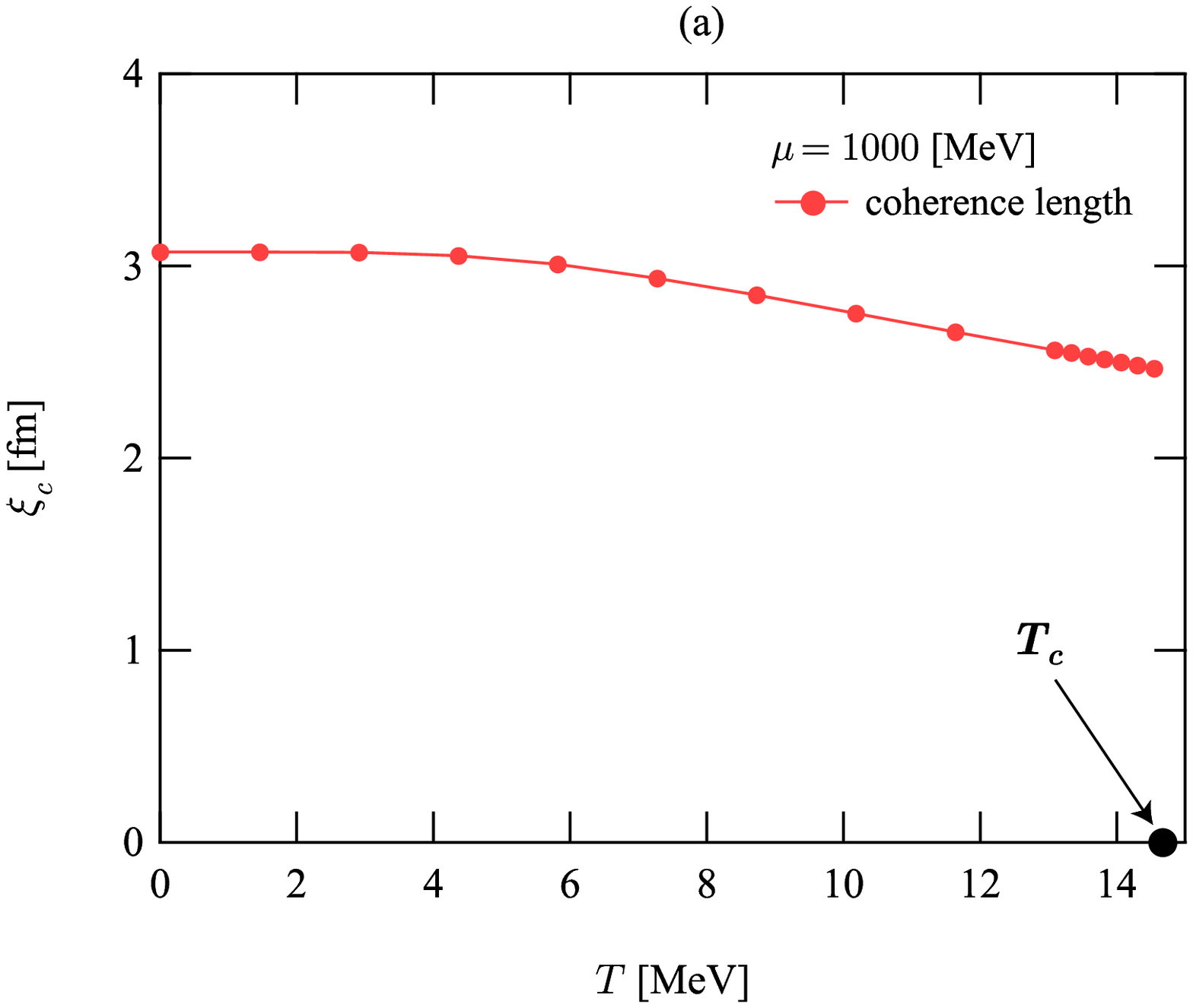}
\includegraphics[width=0.51\textwidth]{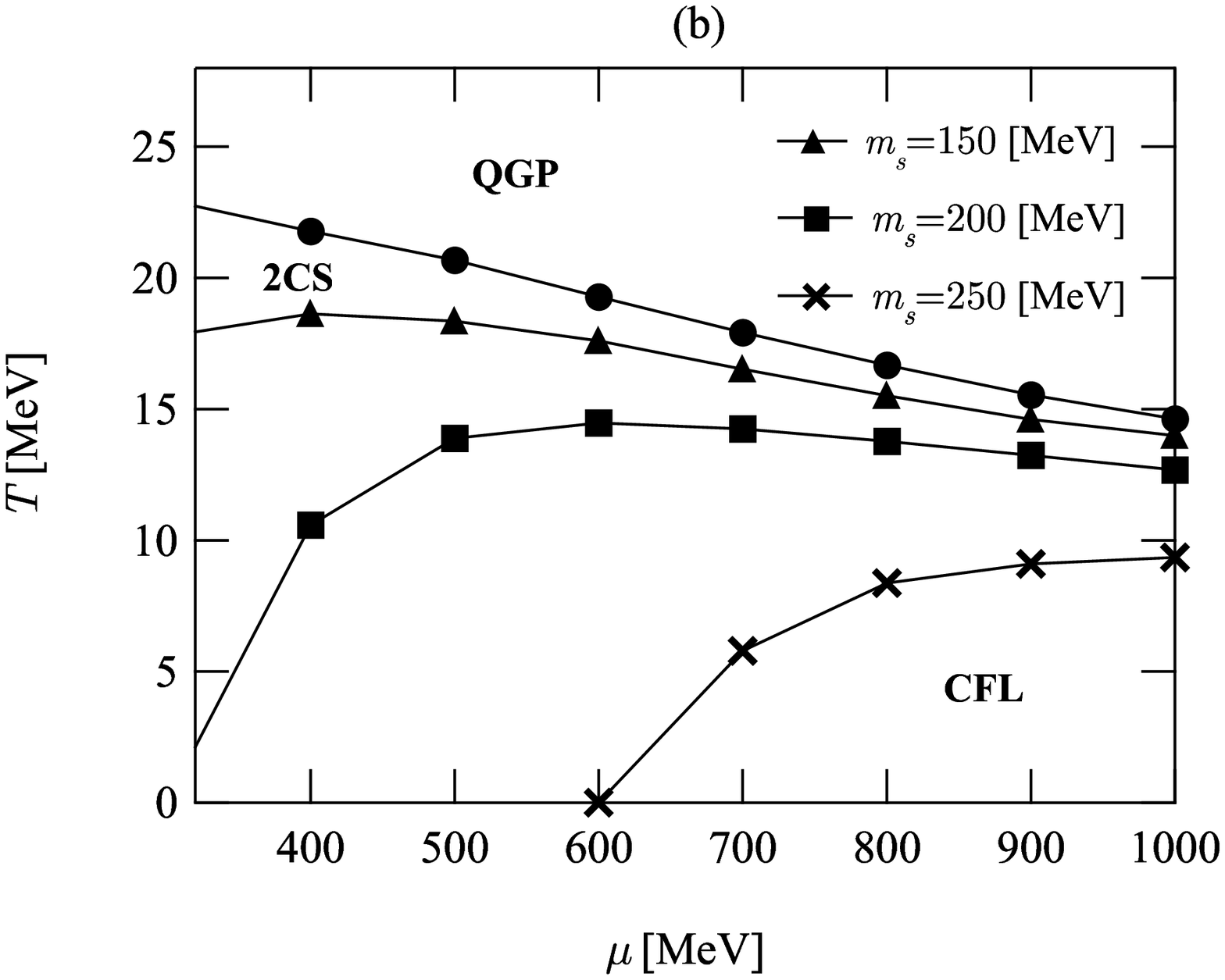}
}
\caption[]{
 (a)~The size of 2SC Cooper pair as a function 
    of temperature at $\mu=1000$MeV.
 (b)~The calculated QCD phase diagram from the improved Schwinger-Dyson
 approach.}
\label{fig:final}  
\end{figure}

We present the numerical solutions for the gap equations
  at $\mu=1000$~MeV corresponding to about 80 times the normal nuclear
  saturation density $\rho_0=0.17~{\rm fm}^{-3}$. 
We display the gap function evaluated on the quasi-quark
  mass shell $\Delta_\pm(p)$ for the 2SC state Fig.~\ref{fig:0}~(a), 
  and the singlet and octet gap functions, $\Delta_{1\pm}(p)$ and
  $\Delta_{8\pm}(p)$, for the CFL state (b). 
The shapes of these gap functions are all similar as functions of
  momentum, except for their magnitudes
  $|\Delta_8(p)|<|\Delta(p)|<|\Delta_1(p)|$. 
In order to see the characteristics of the phase transitions in
  detail, we show the temperature dependence of the gaps at the Fermi
  surface Fig.~\ref{fig:0}~(c) and that of the CJT condensation energies
  for the 2SC and CFL states (d). 
The phase transitions to the QGP phase both from the 2SC state and from
  the CFL phase are of 2nd order, as in the BCS theory with a contact
  interaction:~%
The derivatives of the gaps seem to diverge as
  $T_c$ is approached in (c), and the thermodynamic bulk quantities in
  the superconducting phases connect continuously to those of the
  normal QGP phase (d).
The size of Cooper pair $\xi_c$, the root mean square radius of the
  Cooper pair with wave function $\phi_+$ is displayed in
  Fig.~\ref{fig:final}~(a). 
$\xi_c$ is not affected significantly by changes in the temperature. 
This shows the fact that the Cooper pairs are robust against the thermal
  fluctuation, although the correlation between diquarks diverge 
  as the critical temperature is approached because $\Delta(q)\to0$ as
  $T\to T_c-0$.
This fact also suggests the existence of the incoherent Bose gas
  of tightly bound diquarks just above $T_c$.

Fig.~\ref{fig:final}~(b) shows our result for the QCD phase diagrams for
  $m_s=0,~150,~200$ and $250$~MeV. 
In the chiral limit, there is only one line (black bold dots) dividing
  the $(\mu,T)$ plane into the QGP and CFL phases by 2nd order
  transition. 
The critical temperature is increased as one goes towards low density.
This is due to the strong coupling effects including participation of 
  antiquarks to pairing correlation.
It is surprising that the critical temperatures of transitions to the
  QGP phase from the 2SC and CFL states exactly coincides.
The unlocking critical line on which the CFL state turns into 2SC state
  by 1st order transition appears for finite value of $m_s$, and shifts
  from the triangles to the squares, and to crosses, with the increase
  of $m_s$.
These were estimated using a kinematical criterion for the
  unlocking\cite{unlock_SW,unlock_AR}.
The strength of the 1st order transition becomes weaker as the quark
  density is increased, and the critical end point at which
  the 1st order transition terminates is located at $\mu=\infty$.
If the strange quark mass is lower than $200$ MeV, then chirally broken
  vacuum phase might be continuously connected to the CFL phase without
  phase transition.

\end{document}